\documentclass[twocolumn,showpacs,preprintnumbers,amsmath,amssymb]{revtex4}

\usepackage{graphicx}
\usepackage{dcolumn}
\usepackage{amsfonts,amsmath,amssymb,bm}
\usepackage{epsfig}

\begin{document}

\title{\bf {Transmission spectra of Fe/MgO (001) double-barrier tunnel 
junctions at finite bias}} 
\date{\today} 

\author{J. Peralta-Ramos$^a$}
\email[Corresponding author: ]{peralta@cnea.gov.ar}
\author{A. M. Llois$^{a,b}$}

\affiliation{a) Departamento de F\'isica, Centro At\'omico Constituyentes, Comisi\'on 
Nacional de Energ\'ia At\'omica, Buenos Aires, Argentina\\
b) Departamento de F\'isica, Facultad de Ciencias Exactas y Naturales,
Universidad de Buenos Aires, Buenos Aires, Argentina\\}

\begin{abstract}
In this contribution, we calculate in a self-consistent way the 
ballistic transmission as a function of energy 
of one Fe/MgO (001) single-barrier and one 
double-barrier 
tunnel junction, relating them to their electronic structure. The transmission 
spectra of each kind of junction is calculated at different applied bias voltages. 
We focus on 
the impact that bias has on the resonant tunneling 
mediated by surface and quantum well states. 
The calculations are done in the coherent regime, 
using a combination of density functional theory 
and non-equilibrium Green's functions, as implemented in the {\it ab initio} 
code {\it SMEAGOL}. 

We conclude that, for both kinds of junction, the transmission functions depend 
on the applied bias voltage. In the single-barrier junction, transport 
mediated by resonant Fe minority surface states is rapidly 
destroyed by bias. In the 
double-barrier junction, the appearance of resonant tunneling through 
majority quantum well states is strongly affected by bias. 
\end{abstract}

\maketitle

\section{Introduction}

Magnetic tunnel junctions (MTJs), consisting of a semiconducting barrier 
sandwiched by two ferromagnetic electrodes, are prototype nanoscale systems 
exhibiting spin-dependent electronic transport, 
and are nowadays being intensively investigated due to 
their potential applicability in spintronic devices based on 
the tunneling magnetoresistance effect (TMR). Besides their 
practical importance, MTJs are interesting heterostructures 
on their own since 
they allow us to test our electronic transport theories 
and models, and to understand the complex relationship that  
the electronic, magnetic, and interface structures have 
with novel spin-dependent transport phenomena  
(see Refs. \cite{heil} and references therein). 

One of the main 
challenges in this field is to obtain larger TMR and 
$V_{1/2}$ values (the bias voltage value at which the 
TMR drops to half its value at infinitesimal bias), 
both aspects being critical for device applications \cite{song,mont}. 
As first suggested by Zhang {\it et al} \cite{zhang} 
a decade ago, a possible 
route to accomplish these higher values is to use double-barrier 
MTJs (DBMTJs), in which a metallic slab 
(magnetic or not) is inserted in between the 
semiconducting spacer. In these junctions, whose magnetic and 
transport properties have been measured  
only very recently \cite{noz1,noz2,mont,iovan}, and whose transport 
properties have been calculated mainly using free electron models or  
non-selfconsistently \cite{pet}, 
the in-between metallic slab plays a dual 
role. First, it introduces {\it quantum well states} which can, 
in principle, couple 
to the evanescent states in the spacer, thus producing conductance 
resonances in some spin channel and therefore enhancing the TMR 
\cite{noz2,zhang,iovan,pet,wang}. 
Second, if it is magnetic, 
it introduces a spin-dependent potential energy profile which 
may act as an additional {\it spin filter}, again enhancing  
the TMR \cite{iovan,prbnos,ivnos}.  

In particular, Fe/MgO (001) single- and double-barrier junctions 
are ideal systems to study and good candidates for 
applications, since they show very large TMR values and   
can be nowadays grown epitaxially with 
controlled interfaces. For example, T. Nozaki {\it et al} \cite{noz1,noz2} 
have recently 
measured the TMR of Fe/MgO (001) SBMTJs and identically grown 
DBMTJs as a function of 
the applied voltage and found that the DBMTJs show larger TMR and 
$V_{1/2}$ values. Similar findings have been reported in double-barrier 
junctions of other materials \cite{mont}. 
Although a complete understanding of these features (particularly  
the larger $V_{1/2}$ values) is still lacking, 
we have recently shown by {\it ab initio} calculations on Fe/MgO DBMTJs  
\cite{ivnos} that the spin-filter effect, introduced by the magnetic 
in-between Fe layers, is one 
of their origins. On the other hand, 
A. Iovan {\it et al} \cite{iovan} have recently measured 
extremely large TMR values (an order of magnitude larger than 
those of identically grown MTJs) and conductance oscillations in Fe/MgO DBMTJs, 
originated from resonant tunneling processes mediated by spin-split quantum 
well states. Similar conductance oscillations in Fe/MgO DBMTJs 
have been reported by 
Nozaki {\it et al} as well \cite{noz2}.  
From the theoretical side, Y. Wang {\it et al} \cite{wang} have shown, 
from first principles, that 
the small conductance oscillations 
observed \cite{noz2} in Fe/MgO DBMTJs originate from the majority 
$\Delta_1$-symmetry QWS at the $\Gamma$ point, which 
efficiently couples to a decaying state in the MgO spacer. 
From these 
examples it is clear that DBMTJs are promising candidates for 
spintronic applications, and consequently 
that further theoretical and experimental studies of these 
heterostructures  
are necessary to 
fully understand the role of QWSs and, especially, 
of the bias voltage on their 
spin-dependent transport properties.

With this aim in mind, here we calculate the coherent transmission   
spectra of ideal Fe/MgO (001) SBMTJs and DBMTJs at different values of the applied 
bias voltages. The focus is put on the impact that this bias voltage has on 
the transmission resonances mediated by quantum well states in DBMTJs and 
by surface states in SBMTJs \cite{heil,ivan}. To this end, we employ the 
self-consistent {\it ab initio} {\it SMEAGOL} code 
\cite{smeagol1,smeagol2}, 
which is based on a combination of the {\it SIESTA} 
package \cite{siesta} and the non-equilibrium Green's function 
formalism \cite{daniel}. By extending the work 
of Y. Wang 
{\it et al} \cite{wang} that considered only the 
$\Gamma$ point, we take into account a large 
number of wave vectors inside the  
surface Brillouin Zone. Furthermore, the transport calculations are 
performed self-consistently, taking full account of the 
non-equilibrium electronic population inside the junction induced by a 
finite applied bias voltage to the Fe electrodes.  
We have already presented the 
calculated characteristic 
curves of Fe/MgO (001) SBMTJs and DBMTJs and discussed the different transport 
mechanisms elsewhere \cite{ivnos}, reaching qualitative agreement 
with available experimental results. 
Here, we would like to focus on the impact that 
a bias voltage has on these mechanisms, 
emphasizing the differences between the transmission functions in the  
single- and double-barrier cases. 

\section{Calculation details}

Our SBMTJs consist of 2 monolayers (MLs) of MgO (001), representing 
4~\AA, sandwiched by two {\it semi-infinite} bcc Fe (001) 
electrodes, while our DBMTJs are multilayers of the type 
(MgO)$_2$/Fe$_m$/(MgO)$_2$ (001) sandwiched by 
the same electrodes. The number of in-between Fe monolayers is set equal to $m$=2 MLs, 
representing 2.87 \AA. 
In both cases, the junctions are assumed to be periodic in the {\it x-y} plane, 
being $z$ the transport direction. In this work, we restrict to the parallel magnetic 
configuration of the junctions (hereafter denoted $P$), in which the magnetization vector 
in every magnetic region is parallel to each other. 

In order to account for the charge transfer and to correctly reproduce the band 
offset between Fe and MgO, 
we include in the cell for self-consistent calculations four Fe MLs belonging to the 
electrodes at both sides of the junction.
This in enough to correctly account for charge screening inside the ferromagnet. 
Similar to previous calculations 
\cite{wang,ivan,mc} the lattice constant of the electrodes is fixed to 2.87~\AA~ 
and that of MgO is 
taken to be $\sqrt{2}$ larger. This, together with a 45$^\circ$ rotation of the Fe unit cell, allows  
epitaxial matching between Fe and MgO. 
In this work, the possible appearance of FeO interfacial layers as well as atomic relaxation 
and disorder, are not considered. 

For the electronic structure of the junctions, we use norm-conserving pseudopotentials, 
double-zeta basis set 
for all the angular momenta and the generalized gradient approximation (GGA) \cite{PBE} to the exchange 
and correlation potential. We have thoroughly checked that the band structure and the density of states of 
bulk Fe, bulk MgO and Fe/MgO multilayers, as well as the charge transfer and magnetic moments in the  
last case, are very well reproduced as compared to FP-LAPW results obtained using the highly accurate 
{\it WIEN2k} code \cite{wien}. We obtain a band offset (the difference between the Fermi energy 
$E_\mathrm{F}$ of Fe and the valence band of MgO) of 3.51 eV, in very good agreement with previous 
theoretical \cite{yu} and experimental reports \cite{klaua}. 
As well-known, density functional calculations using semi-local exchange and correlation functionals
underestimate the band gap and ours are not an exception. We obtain a band gap of 5.4~eV 
(as compared to the experimental value of 7.8~eV \cite{klaua,whited}), which agrees well with 
what expected from GGA
\cite{mc,yu}.
 
The ballistic transmission coefficient $T^\sigma (E,V)$ is calculated for each bias and 
it is given by
\begin{equation}
T^\sigma (E,V)=\frac{1}{V_\mathrm{BZ}}\int dk_x dk_y ~T^\sigma (E,V,k_x,k_y)
\label{k}
\end{equation}
where $V_\mathrm{BZ}$ is the area of the surface Brillouin zone orthogonal to the transport 
direction $z$. 
Here we assume that both spin and transverse momentum $k_\parallel$ 
are conserved, an approximation that is
valid for relatively thin epitaxial junctions. 
The {\it $k_\parallel$-resolved} transmission coefficient appearing in 
Eq. (\ref{k}) is calculated 
self-consistently 
from the non-equilibrium Green's functions formalism in the usual way \cite{smeagol1,smeagol2,daniel}. 
It is given by $T=Tr[\Gamma_\mathrm{L}G^r\Gamma_\mathrm{R}G^a]$, where for simplicity we omit the 
spin label $\sigma$. Here, $\Gamma_\mathrm{L,R}$ are the broadening matrices describing the interaction
(thus the finite lifetime) of the scattering region's energy levels due to the 
interaction with the left and right electrodes, 
and $G^{r}$ ($G^a$) is the associated retarded (advanced) Green's function describing 
the one-electron dynamics inside the scattering region. 
The broadening matrices are calculated from the self-energies $\Sigma_{L,R}$ as 
$\Gamma_{L,R}=i(\Sigma_{L,R}-\Sigma^\dag_{L,R})$. These in turn are obtained with 
the semi-analytic method described in reference \cite{smeagol2}. 

The selfconsistent loop consists in starting from an initial  
density matrix of the scattering region (for example, the equilibrium -no bias- density 
matrix), $\rho^{(1)}$, which gives an initial Kohn-Sham Hamiltonian 
$H_A[\rho^{(1)}]$ \cite{siesta}. With this, 
the $G^{r,a}$ 
can be obtained \cite{daniel,smeagol1}, 
since $\Sigma_{L,R}$ are calculated at equilibrium ($V$=0) 
and 
then rigidly shifted in energy.  That is, the voltage-dependent 
selfenergies $\Sigma_{L,R}(E,V)$ are therefore given by 
$\Sigma_{L,R}(E\mp eV/2,V=0)$. The upper (lower) sign in the first 
argument refers to the left (right) electrode.
These retarded/advanced Green's functions are then fed into Keldysh equation for 
the lesser correlation $G^<=G^r\Sigma^<G^a$ \cite{daniel}, from which a new 
density matrix is obtained. In this expression, 
the lesser self-energy $\Sigma^<=\Sigma^<_L+\Sigma^<_R$ is 
obtained from the local equilibrium hypothesis: 
$\Sigma^<_{L,R}=if_{L,R}\Gamma_{L,R}$, where $f$ is the Fermi-Dirac distribution 
function corresponding to the left and right electrodes.  
With this new density matrix obtained via the Keldysh equation, 
the loop starts again, until convergence is achieved. We 
note that the selfconsistent loop must be performed 
at each bias voltage (because the electron population inside the scattering  
region, induced by the electrodes, depends on $V$). 

In our calculations, we use a $8\times8\times8$ k-point mesh in reciprocal space to 
calculate the density matrix 
of the scattering region and a $150\times 150 \times 1$ mesh to evaluate the 
transmission at each bias voltage. 
We have carefully verified that these meshes are sufficient for converging the 
density matrix and the transmission. 

\section{Results and discussion}

Starting with the single-barrier junction, in Fig. 1 we show the transmission spectra at 
zero bias voltage (upper panel) and at $V$=0.1 Volt (lower panel). 
It is seen that, at zero bias, the 
$P$ minority channel presents a broad peak centered around $E_F$+0.1 eV. This conductance 
peak is the signature of a resonance mediated by the Fe-minority surface state located 
at that energy. We note that this surface state resonance is also present in 
a SBMTJ with $n$= 4 MLs. 
At zero bias (or low enough bias), the Fe surface states located at both 
sides of the MgO barrier are approximately 
aligned in energy, and therefore can resonate through it. As 
it can be seen from the lower panel of Fig. 1, this surface state resonance is washed 
out by applying even a small bias voltage of 0.1 Volt. In such a situation, 
due to the applied bias voltage the surface states at each electrode are no longer aligned 
in energy. As a consequence, the resonance condition is lost and the transmission peak 
disappears. Therefore, as has been 
already shown theoretically by Rungger {\it et al} \cite{ivan} in Fe/MgO SBMTJs, 
transport calculations 
in which the transmission function is bias-independent (often taken as the 
zero-bias transmission) overestimate the 
contribution of surface states to the current. 
Here, we see a first example showing the 
importance of taking into account the dependence of the transmission function on the 
applied bias voltage. 

Another interesting feature from Fig. 1 is that the $P$ majority 
transmission is almost independent of bias, in marked constrast to the $P$ minority 
transmission. This difference is a direct consequence of the different transport 
mechanisms that govern the transmission of each of the two $P$ channels. 
The $P$ minority conductance is dominated 
(at least for SBMTJs with thin barriers) by Fe surface states that can 
couple to each other directly through the MgO slab. 
In contrast, the $P$ majority 
transmission is governed by the MgO slowly-decaying 
complex band of symmetry $\Delta_1$, that can couple efficiently to the 
Fe Bloch states of the same symmetry \cite{mc}. The complex bands are smooth functions 
of the energy (just as real bands are), 
and therefore the application of moderate bias voltages has 
no significant influence on them. 
\begin{figure}[htb]
\includegraphics[scale=0.5]{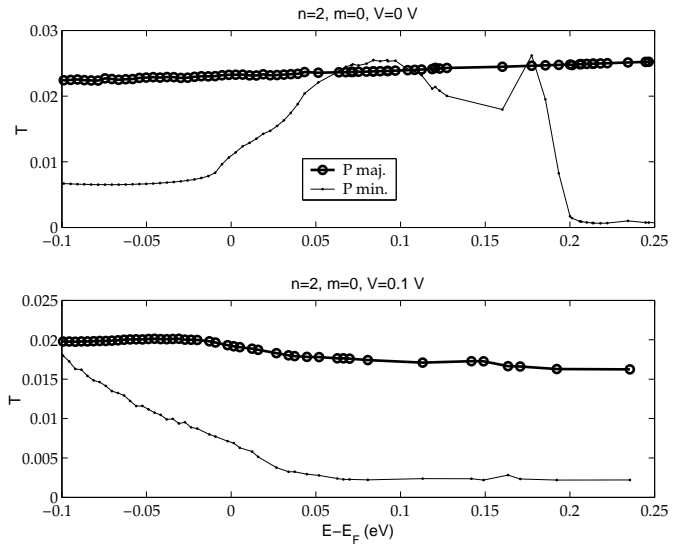}
\caption{Transmission spectra of a single-barrier junction with $n$=2 MLs (4~\AA) 
at zero bias ({\it upper panel}) and at $V$=0.1 Volt ({\it lower panel}).}
\label{Fig1}
\end{figure}

Going over to double-barrier junctions, in Figs. 2-4 we show the transmission spectra 
of the DBMTJ with $n$=2 MLs (4~\AA) and $m$=2 MLs (2.87~\AA), at 
$V$=0.005 Volt, 0.1 Volt and 0.15 Volt, respectively.  At $V$=0.005 Volt  
(Fig. 2) the surface state resonance in the $P$ minority channel is still 
observed, although the peak height is smaller than the one at zero bias shown 
in Fig. 1 for the single-barrier junction. That is to say, the resonant 
mechanism mediated by the Fe surface states can occur even in DBMTJs with 
very thin in-between Fe slabs, as long as the applied bias is small. 
As it happens in the single-barrier junction, 
the application of a bias voltage rapidly destroys this resonance, as it can be 
seen from Figs. 3 and 4 where the transmission resonance is no longer present. 
The origin of this feature is exactly the same as in SBMTJs, namely, the 
misalignment in energy of the Fe surface states at each side of the barrier. 

At $V$=0.005 Volt (Fig. 2) and 
at $V$=0.15 Volt (Fig. 4), the $P$ majority transmissions are very similar to each other, 
being increasing linear functions of the energy. Again, this feature reflects the fact 
that the $P$ majority transmission is esentially governed by the complex bands of MgO. 
But at $V$=0.1 Volt 
(Fig. 3) the $P$ majority transmission is radically different. 
\begin{figure}[htb]
\includegraphics[scale=0.5]{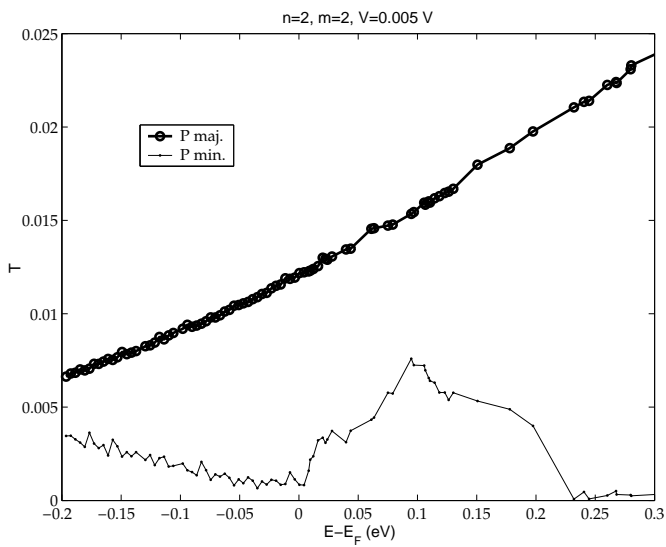}
\caption{Transmission spectra of a double-barrier junction 
with $n$=2 MLs (4~\AA) and $m$=2 MLs (2.87~\AA) 
at $V$=0.005 Volt.}
\label{Fig2}
\end{figure}
\begin{figure}[htb]
\includegraphics[scale=0.5]{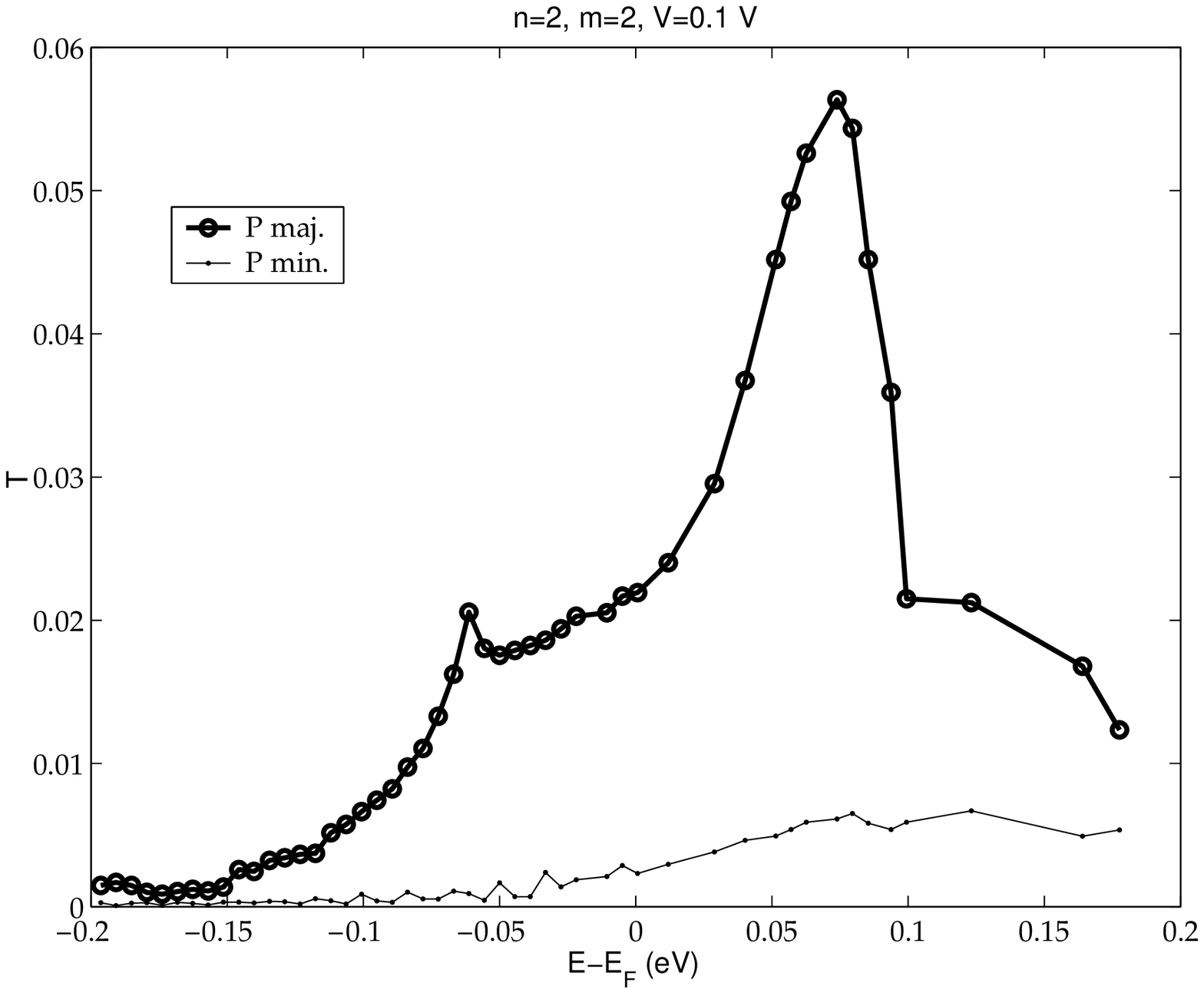}
\caption{Transmission spectra of a double-barrier junction 
with $n$=2 MLs (4~\AA) and $m$=2 MLs (2.87~\AA) 
at $V$=0.1 Volt.}
\label{Fig3}
\end{figure}
\begin{figure}[htb]
\includegraphics[scale=0.5]{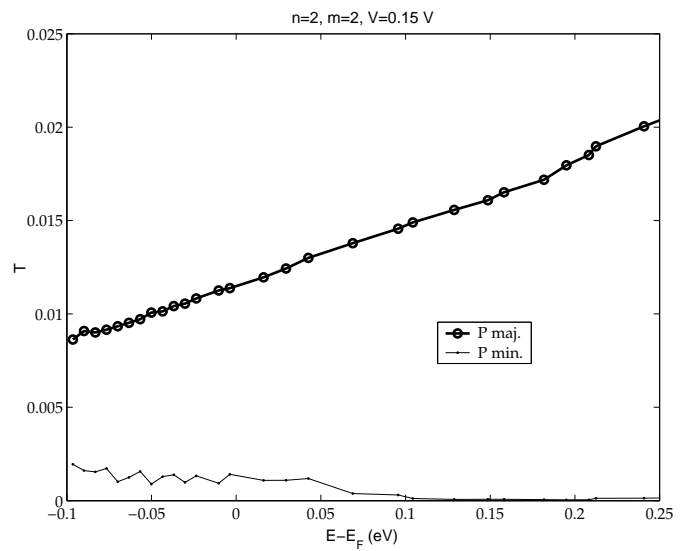}
\caption{Transmission spectra of a double-barrier junction 
with $n$=2 MLs (4~\AA) and $m$=2 MLs (2.87~\AA) 
at $V$=0.15 Volt.}
\label{Fig4}
\end{figure}
It is seen than 
it presents a sharp peak located near $E_F$+0.075 eV. This peak is 
characteristic of resonant tunneling through a quantum well state 
inside the in-between Fe slab. This quantum well state has already 
been shown to be present by 
Wang {\it et al} \cite{wang}, and we have studied its impact on the $I-V$ curves of 
Fe/MgO (001) DBMTJs \cite{ivnos}, reaching 
quantitative agreement with recent experimental results \cite{noz2,iovan}.  
Therefore, from Figs. 2-4 we see that the appearance of 
a transmission resonance mediated by a quantum well state is 
strongly dependent on the applied bias voltage. It is very interesting to 
note that the resonance disappears even with very slight changes in 
the applied bias voltage. A change in bias of 0.05 Volt is 
already sufficient to suppress the quantum well state resonance. 

These results clearly show that, in general, it is 
not possible to assume that the transmission function of tunneling barriers is 
independent of bias, as it is usually done in non-selfconsistent 
calculations \cite{pet}. Both in single- and in double-barrier junctions these  
functions depend on bias, although the dominant transport mechanisms are 
completely different. In particular, in the double-barrier junction that we 
consider, the $P$ 
minority channel behaves, qualitatively, as it does in single-barrier ones, 
being dominated by the resonances mediated by Fe surface states at each side 
of the MgO barrier. The application 
of a bias voltage shifts these surface states in energy, thus forbidding 
their direct coupling to each other. 
On the other hand, the 
$P$ majority channel behaves differently because in double-barrier junctions, 
in addition to transport mediated by the complex bands of MgO, another 
mechanism comes into play, namely, resonant tunneling through quantum 
well states. Our results are just an example of how bias-dependent is this last 
transport mechanism. We are currently investigating these issues in more detail. 
In particular, we are trying to understand the origin of the bias dependence of quantum 
well states resonances and its relation to the potential profile inside the 
scattering region. 

\section{Conclusions}

Using realistic models of electronic structure and self-consistent 
coherent transport calculations based on non-equilibrium Green's functions 
formalism, we 
have obtained the transmission spectra of a single-barrier and of a double-barrier 
Fe/MgO tunnel junction, at different values of the applied bias voltage. We have 
shown that, both in single- and double-barrier junctions, the transmission spectra 
depend on bias. In particular, in single-barrier junctions the application 
of a small bias voltage is already sufficient to suppress resonances between 
Fe minority surface states across the insulating barrier, a fact already 
discussed by Rungger {\it et al} \cite{ivan}. In double-barrier junctions, 
the appearance of resonances mediated by quantum well states in the majority channel 
is strongly dependent on the applied bias. These results bring some insight into 
the phenomena of resonant tunneling through epitaxial magnetic tunnel junctions at 
finite bias, and may be useful in the design and development of future 
spin-electronic devices. 

This work was partially funded by UBACyT-X115, PIP-CONICET 6016, 
PICT 05-33304 and PME 06-117. 
A. M. Llois belongs to CONICET.
(Argentina).


\begin{thebibliography}{99}

\bibitem{heil} C. Heiliger, P. Zahn, B. Y. Yavorsky, and I. Mertig, 
Phys. Rev. B {\bfseries 73}, 214441 (2006); C. Tiusan {\it et al}, 
Phys. Rev. Lett. {\bfseries 93}, 106602 (2004); P. Xu 
{\it et al}, Phys. Rev. B {\bfseries 73}, 180402 (2006); C. 
Heiliger, M. Gradhand, P. Zahn, and I. Mertig, Phys. Rev. Lett. 
{\bfseries 99}, 066804 (2007); J. Peralta-Ramos and 
A. M. Llois, Physica B {\bfseries 398}, 393 (2007)

\bibitem{song} C. Song, X. J. Liu, F. Zeng, and F. Pan, 
Appl. Phys. Lett. {\bfseries 91}, 042106 (2007)

\bibitem{mont} F. Montaigne {\it et al.}, Appl. Phys. Lett. {\bfseries 73}, 
2829 (1998); S. Ohya, P. N. Hai, and M. Tanaka, Appl. Phys. Lett. 
{\bfseries 87}, 012105 (2005); 
J. H. Lee {\it et al.}, 
J. Magn. Magn. Mater. {\bfseries 286}, 138 (2005); 
Z-M. Zeng {\it et al.}, Phys. Rev. B {\bfseries 72}, 054419 (2005)

\bibitem{zhang} X. Zhang, B-Z. Li, G. Sun, and F-C. Pu, 
Phys. Rev. B {\bfseries 56}, 5484 (1997)

\bibitem{noz1} T. Nozaki, A. Hirohata, N. Tezuka, S. Sugimoto, and K. Inomata, 
Appl. Phys. Lett. {\bfseries 86}, 082501 (2005)

\bibitem{noz2} T. Nozaki, N. Tezuka, and K. Inomata, Phys. Rev. Lett. 
{\bfseries 96}, 027208 (2006)

\bibitem{iovan} A. Iovan {\it et al.}, 
arxiv:0705.2375v1 [cond-mat.mes-hall] (16 May 2007)

\bibitem{pet} A. G. Petukhov, A. N. Chantis, and D. O. Demchenko, Phys. Rev. Lett. 
{\bfseries 89}, 107205 (2002); N. Ryzhanova, G. Reiss, F. Kanjouri,
 and A. Vedyayev, 
Phys. Lett. A {\bfseries 329}, 392 (2004); X. Zhang, B.-Z. Li, 
G. Sun, and F.-C. Pu, Phys. Rev. B {\bfseries 56}, 5484 (1997); 
M. Wilczy\'nski, J. Barna\'s, and R. \'Swirkowicz, 
J. Magn. Magn. Mater. {\bfseries 267}, 391 (2003); W.-S. Zhang, 
B.-Z. Li, and Y. Li, Phys. Rev. B {\bfseries 58}, 14959 (1998)

\bibitem{wang} Y. Wang, Z.-Y. Lu, X.-G. Zhang, and X. F. Han, 
Phys. Rev. Lett. {\bfseries 97}, 087210 (2006)

\bibitem{prbnos} J. Peralta-Ramos, and A. M. Llois, Phys. Rev. B 
{\bfseries 73}, 214422 (2006)

\bibitem{ivnos} J. Peralta-Ramos, A. M. Llois, I. Rungger, and S. Sanvito, 
{\it I-V curves of Fe/MgO (001) single- and double-barrier tunnel junctions} 
(in preparation)

\bibitem{ivan} I. Rungger, A. R. Rocha, O. Mryasov, O. Heinonen, 
and S. Sanvito, J. Magn. Magn. Mater. {\bfseries 316}, 481 (2007)

\bibitem{smeagol1} A. R. Rocha, V. M. Garc\'ia-Su\'arez, S. Bailey, 
C.J. Lambert, J. Ferrer, and S. Sanvito, Phys. Rev. B 
{\bfseries 73}, 085414 (2006); A. R. Rocha, V. M. Garc\'ia-Su\'arez, S. Bailey, 
C.J. Lambert, J. Ferrer, and S. Sanvito, Nature Mater. {\bfseries 4}, 335 (2005)

\bibitem{smeagol2} S. Sanvito, C. J. Lambert, J. H. Jefferson, and 
A. M. Bratkovsky, Phys. Rev. B {\bfseries 59}, 11936 (1999)

\bibitem{siesta} J. M. Soler {\it et al.}, J. Phys. Condens. Matter {\bfseries 14}, 2745 (2002)

\bibitem{daniel} S. Datta, {\it Electronic transport in mesoscopic systems}  
(Cambridge University Press, Cambridge, 1999); 
H. Haug y A.-P. Jauho, {\it Quantum kinetics in transport and 
optics of semiconductors} (Springer-Verlag, Berlin, 1996); 
C. Caroli, R. Combescot, P. Nozieres, and D. Saint-James, 
J. Phys. C: Solid St. Phys. {\bfseries 4}, 916 (1971)

\bibitem{mc} J. M. MacLaren, X.-G. Zhang, W. H. Butler, and X. Wang,
Phys. Rev. B {\bfseries 59}, 5470 (1999); W. H. Butler, X.-G. Zhang, 
T. C. Schulthess, and J. M. MacLaren, Phys. Rev. B 
{\bfseries 63}, 054416 (2001)

\bibitem{PBE} J.P.~Perdew, K.~Burke, and M.~Ernzerhof, Phys. Rev. Lett. {\bf 77}, 3865 (1996).

\bibitem{wien} P. Blaha {\it et al}, {\it WIEN2k: An augmented plane wave 
+ local orbitals program for calculating crystal properties} 
(Techn. Universitat Wien, Wien, 2002)

\bibitem{yu} B. D. Yu, and J.-S. Kim, Phys. Rev. B {\bfseries 73}, 
125408 (2006); K. D. Belashchenko, J. Velev, and E. Y. Tsymbal, 
Phys. Rev. B {\bf 72}, 140404(R) (2005)

\bibitem{klaua} M. Klaua {\it et al.}, Phys. Rev. B {\bfseries 64}, 
134411 (2001)

\bibitem{whited} R. C. Whited, C. J. Flaten, and W. C. Walker, 
Solid State Commun. {\bfseries 13}, 1903 (1973)

\end{thebibliography}
\end{document}